\documentclass[prl,twocolumn,showpacs,showkeys,preprintnumbers,amsmath,amssymb,
  nofootinbib]{revtex4}
\usepackage{amsmath}
\usepackage{amssymb}
\usepackage{amsfonts}
\usepackage{eucal}

\newcommand{\be}{\begin{equation}}
\newcommand{\bse}{\begin{subequations}}
\newcommand{\ese}{\end{subequations}}
\newcommand{\bea}{\begin{eqnarray}}
\newcommand{\eea}{\end{eqnarray}}
\newcommand{\ba}{\begin{array}}
\newcommand{\ea}{\end{array}}
\newcommand{\ee}{\end{equation}}

\begin{document}
\begin {center}
\preprint{
           hep-ph/0502150 \cr
           IPM/P-2005/008 \cr
       UCLA/05/TEP/6 \cr
}\end{center}
\vspace*{2mm}

\title{Pulsar Kicks from Majoron Emission}

\author{Yasaman Farzan$^1$, Graciela Gelmini$^2$, and Alexander Kusenko$^2$}

\affiliation{$^1$  Institute for Studies in Theoretical Physics and
  Mathematics (IPM)
P.O. Box 19395-5531, Tehran, Iran\\
$^2$
Department of Physics and Astronomy, University of California, Los Angeles,
  CA 90024-1547, USA
}

\begin{abstract}
We show that Majoron emission from a hot nascent neutron star can
be  anisotropic in the presence of a strong magnetic field.  If
Majorons carry a non-negligible fraction of the supernova energy,
the resulting recoil velocity of a neutron star can explain the
observed velocities of pulsars.

\end{abstract}
\date{\today}
\maketitle

\begin{center}
\section{I. Introduction}
\end{center}\vspace*{2mm}

Pulsar velocities present a long-standing
puzzle~\cite{kusenko_review}.  The distribution of pulsar
velocities is non-gaussian, with an average velocity (250--500)
km/s~\cite{average,15}.  However, as many as 15\% of pulsars have
velocities greater than 1000~km/s~\cite{15}. Pulsars are
magnetized rotating neutron stars born in supernova explosions of
ordinary stars, and so one expects these high velocities to
originate in the supernova explosions. However, a pure
hydrodynamical asymmetry does not seem to be sufficient to account
for such high velocities.  According to advanced 3-dimensional
calculations, pulsar velocities from an asymmetric collapse alone
should be lower than 200~km/s~\cite{3dim}.  Some earlier papers
have claimed somewhat greater velocities, but, by any account, it
seems unlikely that the asymmetries in the collapse could explain
the high-velocity population with speeds in excess of 1000~km/s.

A much greater energy pool is in neutrinos that take away 99\% of
the supernova energy.  Even an  anisotropy as small as a few per
cent in the neutrino emission is sufficient to explain the
observed pulsar velocities. Of course, the neutrinos are produced
in weak processes whose rates depend on the angle between the
neutrino momentum and the electron spin.  In the strong magnetic
field of a pulsar the electrons are polarized, and neutrinos are
produced with a considerable anisotropy.  It was suggested that
the weak interactions alone could lead to an anisotropic flux of
neutrinos and explain the pulsar kicks~\cite{drt}.  However, the
asymmetry is quickly erased by scattering of the neutrinos on
their way out of the neutron star. In fact, one can show that, in
an approximate thermal and chemical equilibrium an anisotropy in
production or scattering amplitudes cannot result in an
anisotropic flux~\cite{eq}.

There are two ways to evade this no-go theorem~\cite{eq}.  One is to
consider an ordinary neutrino outside its neutrinosphere, where it is {\em
not} in thermal equilibrium.  For example, conversions from one neutrino
type to another between their respective neutrinospheres, in the area where
one of them is trapped, but the other one is free-streaming, could explain
the pulsar kicks~\cite{ks96}.  However, present constraints on the neutrino
masses do not allow the resonant neutrino oscillations to take place at
densities around the neutrinospheres, and so this mechanism does not work.

Another possibility is that there is a new particle, whose
interaction with matter is even weaker than that of neutrinos such
that it is produced out of equilibrium.  Then the no-go theorem of
Ref.~\cite{eq} does not apply.  It has been proposed that an
asymmetric emission of  sterile neutrinos could explain the pulsar
kicks~\cite{sterile}.  In this paper we consider a different
mechanism, based on the emission of Majorons from a cooling
newly-formed neutron star.

Majorons, $\Phi$, are massless pseudo-scalar particles
\cite{graciela} which, to a good approximation, have interactions
only with neutrinos described by the Lagrangian  \be {\cal L}_{\rm
int}= \frac{\Phi}{2}(g_{\alpha \beta} \nu_\alpha^T\sigma_2
\nu_\beta +g_{\alpha\beta}^* \nu_\beta^\dagger \sigma_2
\nu_\alpha^*). \label{int} \ee The role of the  Majoron emission
in the supernova cooling process has been studied
extensively~\cite{cooling,mine}. Inside a supernova core neutrinos
have an effective potential given by \be {\cal L}_{\rm
eff}=-\nu_\alpha^\dagger V_{\alpha \beta} \nu_\beta, \ee where
$V_{\alpha \beta}={\rm diag}(V_e,V_\mu,V_\tau)$ and \be
V_e=\sqrt{2} G_F n_B(Y_e+2Y_{\nu_e}-Y_n/2), \label{poteniale} \ee
\be V_\mu=V_\tau=\sqrt{2}G_Fn_B (Y_{\nu_e}-Y_n/2).
\label{potentialmu} \ee Here, $Y_i=(n_i-\bar{n}_i)/n_B$ and $n_B$
is the baryon density. We note that, for the values of the Majoron
couplings we consider, the terms in the potential due to the
Majoron exchange~\cite{dolgov} are negligible; in other words,
$|g_{\alpha \beta}|^2 n_B Y_\nu/T^2\ll V_e, V_\mu$.

Because of the nonzero effective potential, the dispersion relations of
neutrinos and antineutrinos inside the core are different, making processes
such as $\nu \nu \to \Phi$ and $\bar{\nu} \to \nu \Phi$ kinematically
possible.  These processes give rise to a Majoron flux, which can transfer
some energy, $E_\Phi$, from the core. Obviously, $E_\Phi$ cannot be as high
as the total supernova energy, $E_{\rm total}=(1.5-4.5)\times
10^{53}$~erg. This is because neutrinos form supernova 1987A have been
observed, and this observation implies that at least a third of $E_{\rm
total}$ was emitted in neutrinos. Based on this observation, one can derive
strong bounds on the couplings~\cite{cooling,mine}: \be g_{ee}<4 \times
10^{-7} \ \ \ g_{\mu \mu}, g_{\tau \tau}<10^{-6}\ \ .
\label{bounds} \ee
However, the data from SN1987a are
not precise enough to rule out the possibility that $E_\Phi$ was a
non-negligible fraction of $E_{\rm total}$. Let us define
\be x\equiv E_\Phi/E_{\rm total}
\label{x}, \ee
and let us assume that the emission of Majorons is anisotropic,
with an asymmetry $\epsilon$ of a few percent.  Then the overall
anisotropy is $\epsilon x$. If this quantity is of the order of
$10^{-2}$, the anisotropic emission would give the neutron star a
recoil consistent with the observed pulsar velocities. We will
show that the neutron star's magnetic field can cause such an
asymmetry.

Let us examine whether the Majorons are trapped.  Inside a
supernova core, the processes $\Phi \to \nu \nu$ and $\nu \Phi \to
\bar\nu$ are kinematically allowed. Indeed, if the couplings are
very large ($g>10^{-5}$), the Majorons are trapped inside the core
so they cannot transfer a significant amount of energy to the
outside~\cite{cooling}. Thus, the bounds from supernova cooling
exclude only a small window in the coupling constant values. In
this paper,  we will concentrate on the coupling constant values
that saturate the bounds in eq.~(\ref{bounds}). For such small
values of the couplings, the mean free path of $\bar \nu \Phi \to
\nu$ is two orders of magnitude larger than the radius of the
supernova core~\cite{mine}. The Majoron decay length is even
larger.  As a result, one can assume that the Majorons leave the
core without undergoing any  interaction or decay. Also as it is
discussed in \cite{mine}, for the values of coupling satisfying
the upper bounds (\ref{bounds}), the four particle interactions
involving Majorons, such as $\Phi \nu \to \Phi \nu$, $\nu \nu \to
\Phi \Phi$ and etc., are negligible.

 Now let us
assume that there is a uniform strong magnetic field in the core
along the $\hat{z}$-direction: $\vec{B}=|\vec{B}|\hat{z}$. In the
presence of such a magnetic field the medium is polarized
\cite{semikoz}, and the average spin of electrons is \be \langle
\vec{\lambda}_e \rangle= -{e\vec{B} \over 2}\left(3 \over \pi^4
\right)^{1/3} n_e^{-2/3}. \label{pol} \ee
As a result, the effective
potential of neutrinos receives a new contribution, $\delta V$
\cite{semikoz}: \be \delta V=-\sqrt{2} G_F Y_e n_B \langle \lambda_e
\rangle \cos \theta {\rm diag}(3/2,1/2,1/2), \ee
where $\theta$ is the angle between the neutrino momentum and the
direction of polarization. Since the effective potential of the
neutrinos depends on the direction of their momentum, the rates of
the processes $\nu\nu \to \Phi$ and $\bar{\nu}\to \nu \Phi$ will
also depend on the direction. The emission of Majorons produced in
these three-particle processes is strongly correlated with the
direction of the initial neutrinos~\cite{mine}. Therefore, the
Majoron emission will be  anisotropic.

We stress that in all our discussion we neglect the neutrino
magnetic moment, which is very small in the Standard Model with
massive neutrinos. The magnetic field affects the neutrinos only
indirectly, through  polarizing  the electrons in the medium. If
some new physics makes the neutrino magnetic moment
non-negligible, it may have implications for the pulsar
kicks~\cite{voloshin}.

The rest of this paper is organized as follows.  In sect. II, we
will evaluate the momentum  that the process $\nu_e \nu_e \to
\Phi$ can exert on the neutron star in terms of the total energy
transferred to Majorons.  In sect. III, we will perform the same
analysis for the processes $\nu_\mu \nu_\mu \to \Phi$ and $ \bar
\nu_\mu \to \Phi\nu_\mu$.  In sect. IV, we summarize our
conclusions and discuss the effects of a realistic configuration
of the magnetic field, which is probably not a pure dipole.

\vspace{0.3cm}

\section{II. Effects of  $\nu_e \nu_e \to \Phi$}
During the first few seconds after the core collapse, inside the
inner core ($r<10$~km), the electron neutrinos are degenerate:
$\mu_{\nu_e}\sim 100-200$~MeV and $T\sim 10-40$~MeV \cite{model}.
Right after the core bounce $V_e$ is positive, which makes the
process $\nu_e \to \bar{\nu}_e \Phi$ kinematically allowed.
However, after about one second $V_e$ becomes negative and instead
of $\nu_e$-decay, $\nu_e \nu_e \to \Phi$ becomes the source for
the production of $\Phi$. As  discussed in Ref.~\cite{mine}, the
time during which $V_e$ is positive is too short to be important
for energy depletion (or momentum transfer). Thus, we concentrate
on the time when $V_e<0$.

Consider two electron neutrinos with momenta
$$p_1=|\vec{p}_1|(1,\sin \theta_1,0,\cos \theta_1)$$ and
$$p_2=|\vec{p}_2|(1, \sin \theta_2 \cos \phi, \sin \theta_2\sin
\phi, \cos \theta_2).$$ The cross-section of $\nu_e(p_1)
\nu_e(p_2) \to \Phi$ is given by \cite{mine}\be \label{sigma}
\sigma={2 \pi g_{ee}^2\over 4 p_1^2 p_2^2|v_1-v_2|}(p_1+p_2)|2
V_e+\delta V_1+\delta V_2|\delta (\cos \theta_3-\cos\theta_0)\ee
where $\cos \theta_3 =\vec{p}_1\cdot \vec{p}_2/
(|\vec{p}_1||\vec{p}_2|)$  and \be \label{theta0} \cos
\theta_0=1+(p_1+p_2)(2V_e+\delta V_1+\delta V_2)/(p_1 p_2).\ee
Note that  $\delta V_1$ and $\delta V_2$ depend on the directions
of $\vec{p}_1$ and $\vec{p}_2$. Integrating over all possible
momenta of the neutrinos, we find that the neutrinos inside a
volume $dV$ during time $d\tau$, transfer a momentum to  the core
which can be estimated as \be \label{dp} d \vec{P}= {7 \sqrt{2}
\over 24} G_F n_e \langle \vec{\lambda}_e \rangle  {|g_{ee}|^2
\over (2 \pi)^3} (\mu_{\nu_e})^4 d V d \tau.\ee
Of course,
the process $\nu_e \nu_e \to  \Phi$ speeds up the deleptonization
process and, therefore, the duration of the neutrino emission becomes
shorter. However, for $g_{ee}<4\times 10^{-7}$, $\Gamma( \nu_e
\nu_e \to \Phi)\ll \Gamma (ep \to \nu_e n)$ and we expect that the
$\beta$-equilibrium is maintained, and the overall evolution
of the density profiles is similar to the case without
Majoron emission \cite{model}.

Since we do not know the value of $|g_{ee}|$, it is convenient to
write the total momentum transferred to the core in terms of the
energy taken away by Majorons, $E_\Phi=x E_{\rm total}$: \be
\begin{split} \int d \vec{P}&={\sqrt{2} G_Fn_e E_{\rm total} x \over 2 |V_e|}
\langle \vec{\lambda_e} \rangle \cr  &=   -{\sqrt{2} G_F E_{\rm
total} x e \over 4 |V_e|}\left(3 n_e \over \pi^4\right)^{1/3}
|\vec{B}|\hat{z}.
\end{split}
\ee
The value of $|V_e|$ changes with time because of the loss of the
electron lepton number through neutrino and Majoron emission.
Calculating the exact time-dependence of $|V_e|$ is beyond the
scope of this paper. Here we take a typical value of 0.5 eV for
$|V_e|$ to estimate the order of magnitude of the effect.

In order that a star of mass $M_s$ gains a velocity of $v$, the
magnetic field has to be as large as
\be \label{bb}\begin{split} |\vec{B}| &=\left({M_s\over 1.4~
    M_\odot}\right) \left({v\over 500~{\rm km/s}}\right)\left({3 \times
    10^{53}~{\rm erg} \over E_{\rm total}}\right)\cr & \times \left(V_e
  \over 0.5~{\rm eV} \right) \left({ 0.05~{\rm fm}^{-3}\over
    n_e}\right)^{1/3}\left(\frac{0.5}{x}\right) 3\times 10^{16}~{\rm G}.\
\end{split} \ee
Little is known about the magnetic fields in the core of a hot
neutron star at birth.  Observations show that magnetic fields at
the {\em surface} of an average radio pulsar millions of years
after birth are of the order of $10^{12}$~G.  However, some of the
observed neutron stars appear to have surface magnetic fields as
high as $10^{15}$~G~\cite{1015}.  It is reasonable to assume that
the field in the core of a neutron star is stronger than it is on
the surface.  It is also likely that the magnetic field inside a
typical neutron star grows to $\sim10^{16}$~G or higher during the
first seconds after the onset of a supernova explosion due to a
dynamo action~\cite{duncan}.  This field subsequently evolves and
decays during the later stages of neutron star cooling.  An
assumption that all neutron stars have strong interior magnetic
fields at birth is not in contradiction with any of the present
data.

We conclude that, if the Majorons carry away a substantial
fraction of the released energy, they can give the pulsar  high
enough velocity to explain the data.

\section{III. Effects of  $\nu_\mu \nu_\mu\to \Phi$ and $
\bar{\nu}_\mu \to \nu_\mu \Phi$}
 The distributions of
$\stackrel{(-)}{\nu_\mu}$ and $\stackrel{(-)}{\nu_\tau}$ in a
supernova core are thermal; however, the densities of these
neutrinos are substantially lower than that of $\nu_e$:
$\mu_{\nu_\mu}=\mu_{\nu_\tau}=0$ and $T\ll \mu_{\nu_e}$. For the
evolution of a neutron star, $\nu_\mu$ and $\nu_\tau$ are
approximately equivalent. So hereafter we collectively call them
$\nu_\mu$ to avoid repetition. In  a supernova core, $V_\mu$ is
negative and as a result, the two processes $\nu_\mu \nu_\mu \to
\Phi$ and $\bar{\nu}_{\mu} \to \nu_\mu \Phi$ can occur. In analogy
with the $\nu_e \nu_e \to \Phi$ case, one can show that in the
presence of a strong magnetic field a net momentum will be
imparted to the supernova core, given by \be
 \begin{split}  \int d \vec{P}&={\sqrt{2} G_Fn_e E_{\rm total} x \over 6
|V_\mu|} \langle \vec{\lambda_e} \rangle \cr  &=   -{\sqrt{2} G_F
E_{\rm total} x e \over 12 |V_\mu|}\left(3 n_e \over
\pi^4\right)^{1/3} |\vec{B}|\hat{z}.\
\end{split}
\ee Again if $x\stackrel{>}{\sim} {0.1}$ and $|\vec{B}|\sim
10^{16}$~G, neutron stars  can gain  high enough velocities.

\section{IV. Discussions and conclusions}
In this paper we have shown that despite the strong bounds on the
Majoron couplings to neutrinos, an asymmetric emission of Majorons
can explain the high velocities of pulsars, provided that a
substantial fraction of the binding energy of the star is emitted
in the form of Majorons ($E_\Phi/E_{\rm total} \stackrel {>}{\sim}
0.1$).  The asymmetric emission can be caused by a magnetic field
 of  order of $10^{16}$~G in the supernova core. Such high magnetic
fields are quite possible in a supernova core~\cite{duncan}.

The bulk of Majorons are produced deep inside the core, where the structure
of the magnetic field is unknown.  Surface magnetic fields are measured at
much later times, when the neutron star is very cold.  One does not
expect a significant correlation between the field inside the core during
the first seconds of a supernova explosion and the field on the surface
of a cold neutron star that emerges from this explosion.  As a result, we
do not expect a correlation between the pulsar velocity and its observed
magnetic field (the so called $B$--$v$ correlation).

As was suggested by Spruit and Phinney~\cite{rotation}, the
mechanism responsible for the large pulsar velocities can also
cause large angular momenta of pulsars. The emission of
Majorons can give rise to a high angular momentum, provided that the
magnetic field is not rotationally symmetric.  The dynamo
mechanism~\cite{duncan} can generate an off-centered dipole
component if the convection at intermediate depths is faster than
in the center.  The latter is, indeed, likely because the negative
entropy and lepton number gradients necessary for convection can
develop in the outer regions, cooled by the neutrino emission.

 {\bf Acknowledgments} We are grateful to J. Valle for useful discussions.
G.~G. and Y.~F. thank Aspen Center for Physics for hospitality
during their stay, when a part of this work was done.  The work of
Y.~F. was supported in part by DOE grant DE-AC03-76SF0051 and the
Iranian chapter of TWAS at ISMO. The work of G.~G. and A.~K. was
supported by DOE grant DE-FG03-91ER40662 and NASA grants
ATP02-0000-0151 and ATP03-0000-0057.

\end{document}